\documentstyle[aps,pra,floats,epsfig]{revtex}

\newcommand{\be}{\begin{equation}}
\newcommand{\ee}{\end{equation}}
\newcommand{\bea}{\begin{eqnarray}}
\newcommand{\eea}{\end{eqnarray}}
\newcommand{\bt}{\begin{tabular}}
\newcommand{\et}{\end{tabular}}
\newcommand{\ba}{\begin{array}}
\newcommand{\ea}{\end{array}}

\begin{document}



\title{\small{\hfill{
\begin{tabular}{l}
DSF$-$9/2003
\\
hep-th/0304058
\end{tabular}}}
\\ Minimal coupling of the Kalb-Ramond field to a scalar field}

\author{E. Di Grezia and S. Esposito
\thanks{Electronic addresses: digrezia@na.infn.it,
salvatore.esposito@na.infn.it}}

\address{Dipartimento di Scienze Fisiche, Universit\`{a} di Napoli
``Federico II''\\ and \\ Istituto Nazionale di Fisica Nucleare,
Sezione di Napoli \\ Complesso Universitario di Monte S. Angelo,
Via Cinthia, I-80126 Napoli, Italy}

\maketitle

\begin{abstract}
We study the direct interaction of an antisymmetric Kalb-Ramond
field with a scalar particle derived from a gauge principle. The
method outlined in this paper to define a covariant derivative is
applied to a simple model leading to a linear coupling between the
fields. Although no conserved Noether charge exists, a conserved
topological current comes out from the antisymmetry properties of
the Kalb-Ramond field. Some interesting features of this current
are pointed out.

\end{abstract}




\section{Introduction}
\label{s1}

\noindent Space-time non-commutativity is one of the key new ideas
which follows from recent developments in string and matrix theory
\cite{Szabo}. Non-commutativity implies general covariance and,
therefore, it seems likely that a non-commutative Yang-Mills
theory is a good candidate for a unified and potentially
renormalizable theory of the fundamental interactions including
gravity.
\\
Whereas the structure of the space-time becomes non-commutative,
we can describe it, in analogy to quantum phase space, in terms of
the algebra generated by non-commuting coordinates:
\begin{equation}
 [x^{\mu },x^{\nu }] = i\theta ^{\mu \nu }(x),
\end{equation}
with \( \theta ^{\mu \nu } \) an antisymmetric tensor \cite{rino}.
\\
Antisymmetric tensor fields are widely used in string models
\cite{kalb},\cite{cremmer},\cite{nambu} as well as in some
supersymmetric theories. For example, they appear naturally in
$N=2$ extended supersymmetry as auxiliary fields (being the
highest-dimension fields in a supermultiplet) and in the
11-dimensional formulation of $N=8$ extended supergravity in which
they are dynamical fields. Furthermore, in quantum gravity,
antisymmetric tensor fields appear as Lorentz ghost fields
\cite{tow}. It is, therefore, quite important to study the
dynamics of such fields and, especially, their coupling to matter
(scalar or fermion particles).
\\
The best studied example of an antisymmetric tensor field is the
electromagnetic field strength $F_{\mu \nu}$, whose dynamics is
very well known (see, for example, \cite{Jackson}). The coupling
of the electromagnetic field to matter fields proceeds, usually,
through the {\it gauge principle} with the aid of the vector
potential $A_\mu$ (a massless rank-1 field): interaction is
introduced in the theory by requiring local gauge invariance for
the matter fields. In the electromagnetic case, the gauge
transformations for the vector potential are $A_\mu(x) \rightarrow
A_\mu(x) + \partial_\mu \alpha(x)$, where $\alpha(x)$ is an
arbitrary $x-$dependent function, and the interaction with a
charged scalar field $\varphi$ (with charge $e$) is obtained by
replacing the usual derivative $\partial_\mu$ with the covariant
derivative $D_\mu$ in the free field Lagrangian:
\begin{equation}
\partial_\mu \rightarrow D_\mu= \partial_\mu + i e A_\mu ~.
\end{equation}
In fact, with this substitution, the total Lagrangian (free fields
+ interaction) becomes invariant under the local transformation:
\begin{equation}
\varphi(x) \rightarrow e^{-i e\alpha(x)}\varphi(x) ~.
\end{equation}
As a consequence of gauge invariance, by virtue of the Noether
theorem, electric charge conservation is obtained. Note, however,
that such a minimal coupling prescription is not the only possible
one; for example, magnetic moment interaction is described by a
Lagrangian term involving directly the physical field strength
$F_{\mu \nu}$ rather than the gauge-dependent vector potential.
\\
Several papers have appeared \cite{botta} in which the
generalization of the gauge principle to (abelian) rank-2
antisymmetric fields (Kalb-Ramond fields) is studied. However the
interaction of a scalar or fermion particle with a Kalb-Ramond
field usually proceeds through the coupling with the Maxwell field
\cite{Dass}. Indeed, the matter particles interact with the
electromagnetic field coupled to a Kalb-Ramond field, so that only
an indirect interaction is allowed.

 While the coupling between such fields and matter fields can
always be introduced by adding an {\it ad hoc} term in the
complete Lagrangian without invoking a gauge principle, the main
problem with an interaction generated by an (abelian) gauge group
transformation comes from the difficulty to construct a covariant
derivative from rank-2 gauge fields in analogy with the
electromagnetic case. The present work is aimed to further study
such a problem by considering, for simplicity, massless fields.
\\
In view of some applications considered below in the paper, in the
next section we briefly review the dynamics of an antisymmetric
tensor field and point out its substantial equivalence (in the
massless case) with that of a (real) scalar field. Instead in
Section III we give a procedure to couple a scalar particle with a
Kalb-Ramond field through a gauge principle with a linear
coupling; some comments on quadratic coupling are also reported.
Finally, in Section IV, we study the dynamics of the interacting
fields, pointing out some peculiar features, while in the last
section we give our conclusions and outlook.

\section{Dynamics of a free Kalb-Ramond field}
\label{s2}

\noindent Let us consider a Maxwell-like gauge theory where the
role of the vector potential $A_\mu$ is played by an antisymmetric
tensor field $\theta_{\mu\nu}$. Its dynamics is described by the
following Lagrangian:
\begin{equation}\label{2.1}
{\cal L}_\theta = - \frac{1}{12}  H_{\lambda \mu \nu }H ^{\lambda
\mu \nu} ~,
\end{equation}
where the field strength $H_{\lambda \mu \nu}$ is defined by:
\begin{equation}\label{2.2}
H_{\lambda \mu \nu } \equiv \partial _{\lambda }\theta _{\mu \nu
}+\partial _{\mu }\theta _{\nu \lambda }+\partial_{\nu }\theta
_{\lambda \mu} ~.
\end{equation}
The equations of motion for the free $\theta_{\mu\nu}$ field are,
then, similar to the Maxwell equations and read as follows:
\begin{equation}\label{2.3}
  \partial_\mu{H}^{\mu \nu \sigma }=0 ~.
\end{equation}
The Lagrangian in Eq. (\ref{2.1}) is invariant under the gauge
transformations:
\begin{equation}\label{2.4}
  \theta_{\mu\nu}\rightarrow \theta_{\mu\nu}^\prime = \theta_{\mu\nu} + \partial_\mu
\Lambda_\nu -\partial_\nu\Lambda_\mu ~,
\end{equation}
where $\Lambda_\nu$ is an arbitrary vector field. This gauge
freedom can be used to simplify the writing of the field
equations. Indeed, for example, by using the ``generalized''
Lorentz condition $\partial^{\mu} \theta_{\mu \nu} =0$ we find
that $\theta_{\mu \nu}$ satisfies the ordinary wave equation:
\begin{equation}\label{2.5}
\Box \theta_{\mu \nu} = 0 ~.
\end{equation}
It is easy to prove that the degrees of freedom of the
antisymmetric tensor field $\theta_{\mu \nu}$ are just the same of
a scalar field $\theta$. Let us consider the dual vector field (in
the sense of Poincar\'e) $\theta_\mu$ defined by:
\begin{equation}\label{2.6}
\theta_\mu \equiv \frac{1}{6} \epsilon_{\mu \nu \rho \sigma}
H^{\nu \rho \sigma} = \frac{1}{2} \epsilon_{\mu \nu \rho \sigma}
\partial^\nu \theta^{\rho \sigma} ~,
\end{equation}
which satisfies the Bianchi-type identity:
\begin{equation}\label{2.6b}
\partial^\mu \theta_\mu  = 0 ~.
\end{equation}
The equations of motion in (\ref{2.3}) now become:
\begin{equation}\label{2.7}
\partial_\mu \theta_\nu - \partial_\nu \theta_\mu =0 ~,
\end{equation}
pointing out that $\theta_\mu$ has to be the gradient of a scalar
field $\theta$:
\begin{equation}\label{2.8}
 \theta_\mu = \partial_\mu \theta ~.
\end{equation}
With these changes of variable, the Lagrangian describing the
system can be cast in a form similar to that for a scalar field:
\begin{equation}\label{2.9}
{\cal L}_\theta = \frac{1}{2} \partial_\mu \theta
\partial^\mu \theta  ~,
\end{equation}
this showing that the dynamics of a massless antisymmetric tensor
field $\theta_{\mu \nu}$ is completely equivalent (on the
classical level\footnote{For the quantum equivalence see, for
example, Ref. \cite{quantum}}) to that of a massless real scalar
field $\theta$. Note, however, that Eq. (\ref{2.8}) is a direct
consequence of the free field equations of motion (\ref{2.7}), so
that the mentioned equivalence holds true only in the case of
non-interacting fields.

\section{Gauge principle with a linear coupling}
\label{s3} \noindent Let us consider a charged scalar field $\phi$
described by the usual Lagrangian:
\begin{equation}\label{3.1}
{\cal L}_{\phi} = \left( \partial_\mu \phi \right)^\dagger \left(
\partial^\mu \phi \right) + V(\phi),
\end{equation}
where $V(\phi)$ is a given scalar potential including, eventually,
a mass term. In this section we study direct interaction of this
field with a Kalb-Ramond field $\theta_{\mu \nu}$ generated by
possible gauge group transformations leading to linear coupling
with the $\theta_{\mu \nu}$ field. \\
In analogy with the
electromagnetic case, we introduce a covariant derivative $D_\mu$
as follows:
\begin{equation}
iD_\mu = i\partial_\mu + kX_\mu , \label{3.2}
\end{equation}
where $X_\mu$ is a rank-1 tensor to be defined in terms of the
$\theta_{\mu \nu}$ field and $k$ is a suitable coupling
constant.\\
The Lagrangian describing the scalar field $\phi$ interacting with
$\theta_{\mu \nu}$ is, therefore:
\begin{equation}
{\cal L}_{int}= (D^\mu\phi)^\dagger(D_\mu\phi) + V(\phi).
\label{3.3}
\end{equation}
Obviously the Lagrangian in Eq. (\ref{3.3}) must be invariant
under the gauge transformation in Eq. (\ref{2.4}), so that the
transformation properties of $X_\mu$ (and the corresponding ones
for the field $\phi$) are crucial in the identification of $X_\mu$
itself.
\\
Note, however, that the field $X_\mu$ cannot be written as a
gradient of a given function $\alpha(x)$, in order to have a
"genuine" interaction Lagrangian in Eq. (\ref{3.3}). Indeed, let
us assume that:
\begin{equation}
  X_\mu = \partial_\mu \alpha ~,
  \label{3.3b}
\end{equation}
and consider a local phase transformation for the scalar field
$\phi$:
\begin{equation}
  \phi\rightarrow\Phi=\phi e^{ik\alpha} ~.
  \label{3.3c}
\end{equation}
Substitution into Eq. (\ref{3.3}) immediately leads to:
\begin{equation}
{\cal L}_{int}= (D^\mu\phi)^\dagger(D_\mu\phi) + V(\phi)
=(\partial^\mu\Phi)^\dagger\partial_\mu\Phi) + V(\phi),
\label{3.3d}
\end{equation}
and since $\Phi$ and $\phi$ represent the same physical object, we
conclude that the interaction introduced by means of the $X_\mu$
field in Eq. (\ref{3.3b}) is fictitious, because it can be
re-absorbed by a phase redefinition of the scalar field.
  \\
 Starting from the
rank-2 tensor $\theta_{\mu \nu}$, the simplest linear choice for
$X_\mu$ is the following:
\begin{equation}\label{3.4}
  X_\mu = \partial^\nu \theta_{\mu \nu} ~.
\end{equation}
Such a definition for $X_\mu$ is, however, useful only when the
"generalized" Lorentz condition is not fulfilled, since in this
case $X_\mu$ is identically zero and no physical interaction
appears. Disregarding this gauge choice, the gauge group for the
scalar field $\phi$ is easily obtained. Indeed, by applying the
gauge transformation in Eq. (\ref{2.4}) to the Lagrangian in
(\ref{3.3}), we find that it remains unchanged if the scalar field
$\phi$ transforms as follows:
\begin{equation}
  \phi\rightarrow\phi '=\phi e^{ik\eta} ~,
  \label{3.5}
\end{equation}
with $\eta=\partial^\nu\Lambda_\nu$. Note that the gauge freedom
for the $\theta_{\mu \nu}$ field is not altered if in  Eq.
(\ref{2.4}) we choose a divergence-less gauge function
$\Lambda_\nu$. In this case $\eta$ is zero and  Eq. (\ref{3.5})
reduces to the unit transformation, so that the gauge group
underlying the choice in Eq. (\ref{3.4}), when it is applicable,
acts as the identity on the scalar field $\phi$. As a consequence,
by means of the Noether theorem,
no conserved charge comes out. \\
The only alternative for $X_\mu$, which is linear in the
$\theta_{\mu \nu}$ field, is the following
\begin{equation}
X_\mu=\theta_\mu=\frac{1}{2}\epsilon_{\mu\nu\rho\sigma}\partial^\nu\theta_{\rho\sigma}.
\label{3.6}
\end{equation}
Now we have no limitation on the gauge condition to be used and is
a simple task to show that the gauge group associated to the
scalar field is, again, the identity. Indeed, it is easy to
convince ourselves that $X_\mu$ in Eq. (\ref{3.6}) is invariant
under the gauge transformation in Eq. (\ref{2.4}), and thus the
Lagrangian in Eq. (\ref{3.3}) is automatically invariant when
$\phi$ is unchanged\footnote{More in general, we can allow a phase
transformation $ \phi\rightarrow\phi '=\phi e^{i\beta}$, with a
constant $\beta$.}:
\begin{equation}
  \phi\rightarrow\phi '=\phi ~.
  \label{3.7}
\end{equation}
We remark that the choice (\ref{3.6}) for $X_\mu$ is not a trivial
one, since $X_\mu$ does not satisfy the constraint (\ref{3.3b}).
In fact, as pointed out in the previous section, the dual field
$\theta_\mu$ can be written as the gradient of $\theta$ (see Eq.
(\ref{3.3})) only for a massless non-interacting Kalb-Ramond
field, which is not the present case. It is remarkable that in the
limit of no interaction the $X_\mu$ field becomes unphysical.
\\
A final remark on the physical dimensions of the coupling constant
is in order. We immediately see that, for the cases considered
above, $k$ has the dimensions of the inverse of a mass, so that
the corresponding theory is non-renormalizable. We point out that
such a property is a peculiar feature of Quantum Gravity, which is
a natural framework, however, for the theory developed here.

\subsection{Non-linear coupling}
\noindent For completeness,we briefly discuss some particular
direct interactions between a scalar field and a Kalb-Ramond field
involving
non-linear terms (which are quadratic in the Kalb-Ramond field). \\
First of all we note that $n$-linear ($n>1$) terms in $\theta_{\mu
\nu}$ appearing in the covariant derivative have to be
gauge-invariant in order to assure the invariance of the
Lagrangian (neglecting terms which are total divergences). Indeed,
when considering the $X_\mu$ field where, for example, non
gauge-invariant bilinear terms in $\theta_{\mu \nu}$ appears, we
recognize that, under a gauge transformation (\ref{2.4}), two
derivatives of two gauge ($4$-vectors) functions $\Lambda_\mu$
come out. In general, the corresponding gauge terms in
$(D^\mu\phi)^\dagger(D_\mu\phi)$ cannot be absorbed by a phase
transformation (containing only one scalar function) of the scalar
field $\phi$.
\\
Therefore, simple choices for $X_\mu$ involve only the
gauge-invariant field $H_{\mu\nu\rho}$ and its dual $\theta_\mu$
and, as a consequence, the gauge group is the identity.
\\
Some example of quadratic interactions are as follows:
\begin{eqnarray}
&& X_\mu = \theta_{\nu}\partial^\nu\theta_{\mu} \label{Q1},\\
&& X_\mu = \epsilon^{\alpha\beta\gamma\delta}\partial^\eta
H_{\mu\alpha\beta}H_{\gamma\delta\eta},
\label{Q2}\\
&& X_\mu =\epsilon_{\mu\nu\alpha\beta}\partial^\nu
H^{\alpha\beta\gamma}\theta_\gamma . \label{Q3}
\end{eqnarray}
Note that in the case considered the Eq. (\ref{Q2}) (and for
similar terms), $X_\mu$ vanishes only when $H_{\gamma\delta\eta}$
satisfies its equation of motion in absence of interaction Eq.
(\ref{2.3}). Finally, we point out that allowing quadratic terms
for $X_\mu$ as those in Eqs. (\ref{Q1})-(\ref{Q3}) (and similar
ones), the coupling constant $k$ in the covariant derivative must
have the dimensions of the inverse of the $4$th power of a mass,
so that the non-renormalizability of the theory is greatly
worsened with respect to the linear coupling case.

\section{Interacting field dynamics.}
\noindent The dynamics of a charged scalar field interacting
directly with a Kalb-Ramond field is described by the following
Lagrangian:
\begin{equation}\label{4.1}
{\cal L}= \left( D_\mu \phi \right)^\dagger \left( D^\mu \phi
\right) -m^2\phi^\dagger\phi -
\frac{1}{12}H_{\mu\nu\rho}H^{\mu\nu\rho} ,
\end{equation}
where the covariant derivative is written as in Eq. (\ref{3.2})
and, for simplicity, we have neglected a scalar potential term. By
expliciting the $X_\mu$ term, we can rewrite Eq. (\ref{4.1}) as:

\begin{equation}\label{4.2}
{\cal L}= \left( \partial_\mu \phi \right)^\dagger \left(
\partial^\mu \phi \right) -(m^2- k^2X^2)\phi^\dagger\phi -
ikX_\mu[\phi \partial^\mu \phi ^\dagger - \phi^\dagger
\partial^\mu \phi ]- \frac{1}{12}H_{\mu\nu\rho}H^{\mu\nu\rho},
\end{equation}
where the second term in the sum accounts for the effective mass
acquired by the scalar particle interacting with the Kalb-Ramond
field, while the following term describes this interaction.
\\The Euler-Lagrange equation for $\phi$ immediately follows:
\begin{equation}\label{4.3}
\partial_\mu \partial^\mu \phi + (m^2- k^2X^2)\phi = 2ikX_\mu \partial^\mu \phi
+ ik \phi  \partial^\mu X_\mu.
\end{equation}
Note that, in the linear coupling case, the last term vanishes,
since $\partial^\mu X_\mu =0$.\\
Instead, the equations of motion for the interacting Kalb-Ramond
field are:
\begin{equation}\label{4.4}
\partial^\sigma H_{\sigma\mu\nu}= J_{\mu\nu},
\end{equation}
with
\begin{eqnarray}\label{4.5}
J_{\mu\nu} = & 2 &\left[[ -ik(\phi \partial^\mu \phi ^\dagger -
\phi^\dagger
\partial^\mu \phi) + 2k^2|\phi|^2X_\sigma]\frac{\partial
X^\sigma}{\partial(\partial_\rho\theta_{\mu\nu})}\right]
\nonumber \\
&-& [-ik(\phi \partial_\sigma\mu \phi ^\dagger - \phi^\dagger
\partial_\sigma \phi) + 2k^2|\phi|^2X_\sigma]\frac{\partial
X^\sigma}{\partial X^\sigma\theta_{\mu\nu}}.
\end{eqnarray}
Given the symmetry properties of the $\theta_{\mu\nu}$ field, it
follows that the current $J_{\mu\nu}$ is antisymmetric while
exchanging the indices $\mu$ and $\nu$:
\begin{equation}\label{4.6}
J_{\mu\nu}=-J_{\nu\mu}.
\end{equation}
Moreover, from the field equations (\ref{4.4}) we deduce that,
despite of the explicit form of $X_\mu$, $J_{\mu\nu}$ is a
conserved current:
\begin{equation}\label{4.7}
\partial^\mu J_{\mu\nu}=0.
\end{equation}
Note that such a property follows, again, from the antisymmetry of
$H_{\sigma\mu\nu}$.
\\Let us now consider the explicit interesting case in which
$X_\mu$ is given by Eq. (\ref{3.6}). After some algebra we obtain
the following expression for the current:
\begin{equation}\label{4.8}
J_{\mu\nu}=\partial^\sigma T_{\sigma\mu\nu}. \label{34}
\end{equation}
with
\begin{equation}\label{4.9}
T_{\sigma\mu\nu}=-2(ik\epsilon_{\sigma\rho\mu\nu}\phi
^\dagger\partial^\rho\phi + k^2|\phi|^2 H_{\sigma\mu\nu}).
\label{35}
\end{equation}
It is remarkable that, in the case considered, the current
$J_{\mu\nu}$ is the gradient of an antisymmetric rank-3 tensor
$T_{\sigma\mu\nu}$. Indeed, as a consequence of this, we have that
Eq. (\ref{4.7}) holds independently of the equations of motion (by
taking the divergence of Eq. (\ref{4.8}) we obtain a vanishing
R.H.S due to the symmetry properties of $T_{\sigma\mu\nu}$), so
that $J_{\mu\nu}$ is a conserved topological current. Moreover on
substituting Eq. (\ref{4.8}) into Eq. (\ref{4.4}) we find that
\begin{equation}\label{4.10}
\partial^\mu \tilde{H}_{\mu\rho\sigma}=0.
\end{equation}
with
\begin{equation}\label{4.11}
\tilde{H}_{\mu\rho\sigma}= H_{\mu\rho\sigma}- T_{\mu\rho\sigma}.
\end{equation}
i.e. the novel field $\tilde{H}_{\mu\rho\sigma}$ can be
interpreted as a "dressed" Kalb-Ramond field satisfying the free
field equation (\ref{4.10}).

\subsection{The symmetry group of the topological current}
Let us now turn back to the topological current $J_{\mu\nu}$ in
Eqs (\ref{34}) and (\ref{35}) and write it as sum of a term
$J_{\mu\nu}^0$, which does not depend explicitly on the
Kalb-Ramond field, and a term $J_{\mu\nu}^1$ which vanishes for
vanishing $H_{\mu\nu\rho}$:
\begin{eqnarray}
J_{\mu\nu} & = & -2 k ( J_{\mu\nu}^0 + k J_{\mu\nu}^1)\\
J_{\mu\nu}^0 &= &\epsilon_{\alpha\beta\mu\nu} (\partial^\alpha
\varphi^\dag)(\partial^\alpha \varphi),\\
J_{\mu\nu}^1 & =& \partial^\alpha (|\varphi|^2 H_{\alpha\mu\nu}.
\end{eqnarray}
The "free" current $J_{\mu\nu}^0$ remains invariant if we perform
a rotation with an imaginary angle $i\alpha$ of the fields
$\varphi,\varphi^*$\footnote{More in general, the current
$J_{\mu\nu}^0$ is left unchanged if the fields $\varphi,\varphi^*$
undergo the following linear transformation:
\begin{equation} \left(
\begin{array}{l}
 \varphi \\
 \varphi^*
\end{array} \right)\rightarrow
\left(\begin{array}{l}
 \varphi^\prime \\
 \varphi^{* \prime}
\end{array} \right) =\left(
\begin{array}{ll}
  a & b  \\
  b^* & a^*
\end{array} \right)
\left(\begin{array}{l}
 \varphi \\
 \varphi^*
\end{array} \right)
+\left(\begin{array}{l}
 c \\
 c^*
\end{array} \right)
\label{T5}
\end{equation}
with $a,b,c$ three arbitrary complex numbers satisfying the
constraint $|a|^2 -|b|^2 = 1$. Note that, since the determinant of
the transformation matrix is equal to $1$, the group of
transformations defined in (\ref{T5}) is that of {\it
equiaffinities} in the complex plane, which is a subgroup of
$SL(2,C).$} :
\begin{equation} \left(
\begin{array}{l}
 \varphi \\
 \varphi^*
\end{array} \right)\rightarrow
\left(\begin{array}{l}
 \varphi^\prime \\
 \varphi^{* \prime}
\end{array} \right) =\left(
\begin{array}{rr}
  \cos{i\alpha} & \sin{i\alpha}  \\
  -\sin{i\alpha} & \cos{i\alpha}
\end{array} \right)
\left(\begin{array}{l}
 \varphi \\
 \varphi^*
\end{array} \right)
\end{equation}

However the term $J_{\mu\nu}^1$, and hence the whole current
$J_{\mu\nu}$, is not invariant under the group defined above, so
that the corresponding topological symmetry can be only viewed as
an approximate invariance of the conserved current in the limit of
a small coupling constant (the $J_{\mu\nu}^1$-term is quadratic in
$k$). The general transformation for the scalar field $\varphi$
leaving invariant the whole current $J_{\mu\nu}$ is the following:
\begin{equation}
\varphi\rightarrow \varphi '= \varphi e^{i(\alpha|\varphi|^2 +
\beta)}, \label{T6}
\end{equation}
with $\alpha, \beta$ two real parameters, and, in the limit
$\alpha =0$, we recover the simple global phase transformation.
However, since Eq. (\ref{T6}) is highly non-linear for $\alpha
\neq 0$, it is not evident the physical meaning corresponding to
the transformation in Eq. (\ref{T6}).

\section{Conclusions}
In this paper we have introduced a direct interaction between a
scalar particle and  a Kalb-Ramond field by defining a covariant
derivative $D_\mu$, where an appropriate auxiliary vector field
$X_\mu$ (depending on the Kalb-Ramond field) appears. Several
possible choices for $X_\mu$ have been studied, leading to linear
or quadratic coupling with the scalar field. In the simple, viable
models considered here, the gauge group underlying the theory is
the identity, so that no conserved Noether charge exists. However,
due to the antisymmetry properties of the Kalb-Ramond field, a
conserved (antisymmetric) topological current arises in the
simplest model, which appears in the equations of motion for
$H_{\mu\nu\rho}$. Since this current is a divergence of a rank-3
antisymmetric tensor, it is possible to define  a "dressed"
Kalb-Ramond field strength, obeying the free field equations,
which describes the dynamics of the interacting field.\\
Some possible applications of our results are concerned with the
theory of gravity, where a zero mass Kalb-Ramond field is the
source of torsion in Einstein-Cartan theory \cite{22}. Moreover,
in recent years, it has been pointed out that the presence of
Kalb-Ramond fields in the background space-time leads to several
interesting astrophysical and cosmological phenomena like cosmic
optical activity and neutrino helicity flip \cite{gr}. This
motivates the study of some important problems related to the
standard Friedman-Robertson-Walker cosmological model in light of
Kalb-Ramond cosmology \cite{cos}, \cite{rino} in an inflationary
framework, where the coupling to a scalar field is crucial.

\acknowledgments

We are indebted with Dr. G. Fiore for his kind cooperation and
many interesting suggestions. Useful discussions with Dr.s R.
Marotta, O. Pisanti and Prof. G. Miele have been appreciated as
well.

\end{document}